\documentclass[aps,amsmath,amssymb,twocolumn,prb,longbibliography]{revtex4-2}
\usepackage{graphicx} 
\usepackage{amsfonts,amsmath,amssymb}
\usepackage{amsthm}
\usepackage{dcolumn}
\usepackage{dsfont,bm}
\usepackage[colorlinks=true,linkcolor=blue,pagecolor=blue,filecolor=blue,menucolor=blue,urlcolor=blue,citecolor=blue,anchorcolor=blue]{hyperref}
\usepackage{xcolor}
\usepackage{soul} 
\usepackage{amsbsy}
\usepackage{float}

\usepackage{bm}

\begin{document}

\title{Fine structure of the cyclotron resonance in heterobilayers of proximitized graphene and transition metal  dichalcogenides}

\author{M. A. Rakitskii$^{1}$}
\author{K. S. Denisov$^{2,1}$}
\email{denisokonstantin@gmail.com}
\author{N. S. Averkiev$^{1}$}

\affiliation{$^{1}$ Ioffe Institute, 194021, St. Petersburg, Russia}
\affiliation{$^{2}$ Department of Physics, University at Buffalo, State University of New York, NY 14260, Buffalo, USA}

\begin{abstract}

A monolayer graphene and its Dirac electrons can be equipped with an enhanced spin-orbit coupling (SOC) when proximitized by other van der Waals (vdW) materials, such as transition metal dichalcogenides (TMDs). In this work we analyze the features of the cyclotron resonance (CR) absorption at quantizing magnetic fields emerging in the presence of proximity-induced spin interactions, including the spin-pseudospin Rashba coupling. We evaluate the spin-textured wave functions of the Landau levels and calculate the absorption spectrum paying special attention to its spin proximity induced modifications. We reveal the formation of a fine double-peak structure of the main interband CR transitions, as well as the presence of additional spin-flip absorption, the combined cyclotron resonance (CCR), centered at different resonant frequencies. The selection rules for CCRs are identified and complemented by the perturbation theory analysis. We also discuss the polarization dependence of the absorption and the proximity-induced emerging magneto-optical responses. Our theory explains the effect of proximity-induced spin interactions for Dirac electrons cyclotron resonance and points out at its experimental verifications.

\end{abstract}

\date{\today}
\maketitle

While a pristine monolayer graphene has insignificant spin–orbit coupling (SOC)~\cite{vzutic2019proximitized,min2006intrinsic,gmitra2009band,castro2009electronic,sichau2019resonance}, 
when proximitized by other materials, such as transition metal dichalcogenides (TMDs)~\cite{gmitra2015graphene,gmitra2016graphene,gmitra2017proximity,garcia2018spin,rao2023ballistic,masseroni2024spin}, magnetic layers~\cite{yang2013proximity,lazic2016effective,asshoff2017magnetoresistance,vzutic2019proximitized,ghiasi2021electrical,zollner2023strong,zollner2022proximity,zollner2022engineering,muniz2024rashba,xu2018spin}, or heavier elements~\cite{marchenko2012giant,sierra2025room}, 
it acquires considerable spin splitting of electronic bands, 
carrying great potential for 
SOC-driven physics of Dirac electrons~\cite{han2014graphene,avsar2020colloquium,sierra2021van,zhao2025novel}. 
Extensive developments in van der Waals (vdW) materials and heterostructures~\cite{novoselov20162d,liu2016van,sierra2021van} have recently led to a revision of 
the feasibility of spintronics in graphene~\cite{han2014graphene,dery2011nanospintronics,wen2016experimental,avsar2020colloquium,sierra2021van,zhao2025novel}:
Graphene-based vdW structures 
have demonstrated 
efficient spin–charge interconversion via direct and inverse spin-galvanic effects~\cite{offidani2017optimal,ghiasi2019charge,benitez2020SHEISGE,galceran2021CSI,ontoso2023unconventional,veneri2022twist}, 
spin-current generation from a finite spin Hall effect~\cite{garcia2017SHE,safeer2019MoS2}, 
and anisotropic spin dynamics~\cite{cummings2017giant,ghiasi2017large,benitez2018anisotropy,zihlmann2018large,offidani2018microscopic,ingla2021electrical,sierra2025room}. 
Adding a magnetic proximity effect~\cite{yang2013proximity,lazic2014graphene,lazic2016effective,asshoff2017magnetoresistance,xu2018spin,vzutic2019proximitized,ghiasi2021electrical,zollner2023strong,zollner2022proximity,zollner2022engineering}, or finite twisting in a multilayer~\cite{david2019induced,naimer2021twist,lee2022charge,veneri2022twist,yang2024twist,perkins2024spin}, 
both important for realizing topological band structures and quantized Hall conductance~\cite{qiao2010quantum,qiao2011two,wang2015proximity,gmitra2016graphene,hogl2020quantum,chang2023colloquium,chou2024topological,sha2024observation}, 
further highlights the great versatility 
in tuning the spin-dependent properties of a proximitized graphene.

\begin{figure}[b]
    \centering
    \includegraphics[width=.8\linewidth]{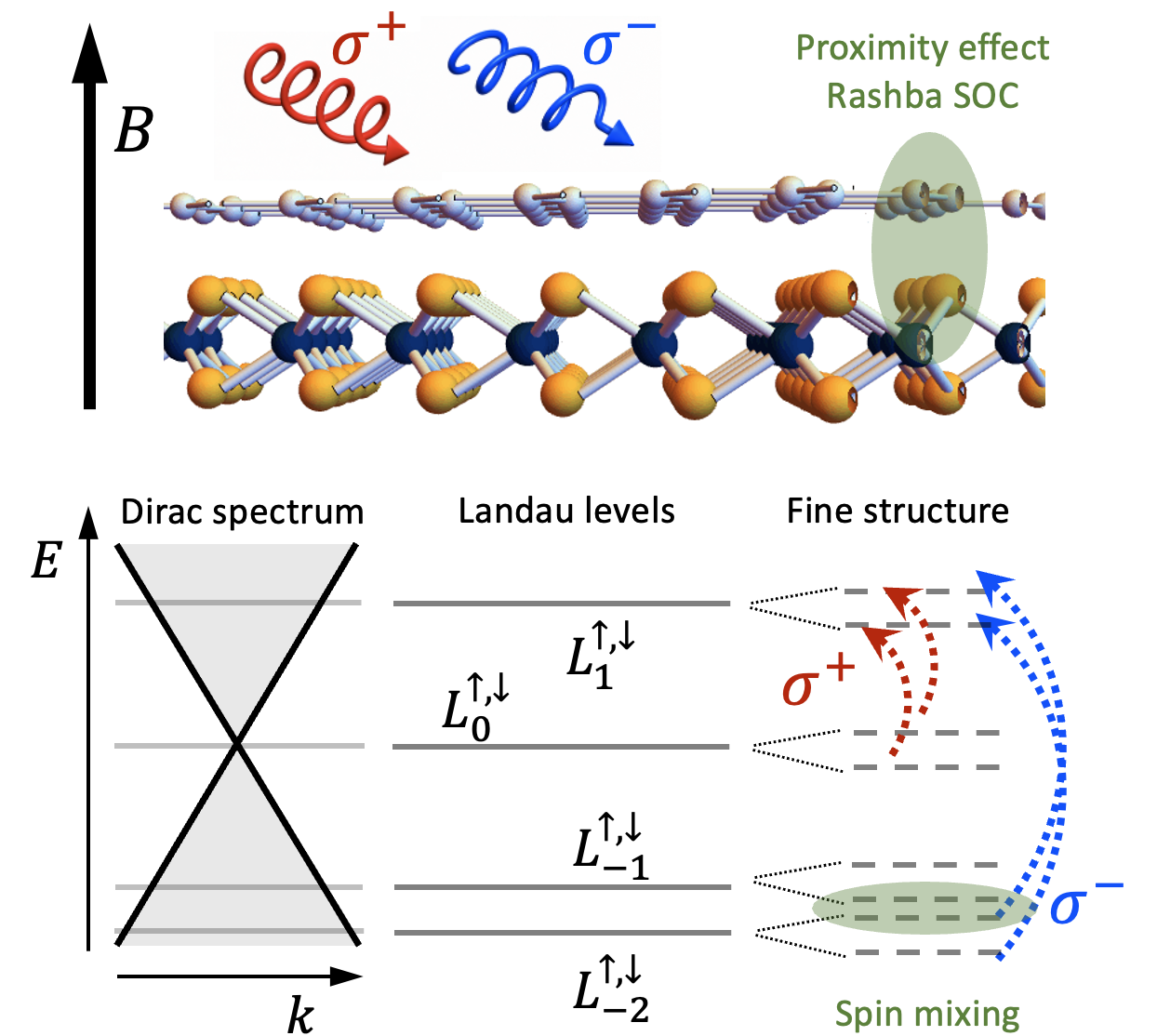}
    \caption{Cyclotron resonance in a graphene/TMDs heterobilayer and associated transitions between Landau levels, $L_n^s$.}
    \label{fig:F0}
\end{figure}

In this paper, we focus on unexplored resonant high-frequency properties of proximitized graphene structures at quantizing magnetic fields, see Fig.~\ref{fig:F0}. Namely, we study the cyclotron resonance (CR) and the combined cyclotron resonance (CCR)~\cite{rashba1960properties} - SOC-induced spin-flip transitions between Landau levels (LLs) allowed in the electric-dipole approximation (EDA), in monolayer graphene with proximity-induced SOC and Rashba coupling~\cite{garcia2018spin,benitez2018anisotropy,safeer2019MoS2,zollner2022proximity,muniz2024rashba}.
The cyclotron resonance spectroscopy and its mid-infrared range (mid-IR) absorption is an alternative method for elucidating the proximity-induced SOC in graphene, which can potentially diversify the results obtained from the previously considered spin relaxation~\cite{cummings2017giant,ghiasi2017large,benitez2018anisotropy,zihlmann2018large,offidani2018microscopic,ingla2021electrical,sierra2025room}, spin-charge interconversion~\cite{offidani2017optimal,garcia2017SHE,safeer2019MoS2,ghiasi2019charge,benitez2020SHEISGE,galceran2021CSI,ontoso2023unconventional,veneri2022twist}, and other approaches~\cite{rao2023ballistic,sun2023determining}. Moreover, in contrast to CCR in the case of a  parabolic spectrum—originally studied by Rashba~\cite{rashba1960properties} and further addressed in bulk~\cite{Bell1962:PRL,McCombe1967:PRL} and nanostructured zincblende semiconductors~\cite{Dyakonov,Stier2023:PRB}—the properties of CCR in graphene-based structures have not been fully elucidated, calling for theoretical analysis.

The cyclotron resonance (CR) in graphene has a number of important features~\cite{sadowski2006landau,gusynin2007anomalous,jiang2007infrared,deacon2007cyclotron,abergel2007optical,orlita2010dirac,orlita2011carrier,scharf2013magneto}. In contrast to a parabolic spectrum, where LLs have an equidistant energy separation $E_n=\hbar\omega_c(n+1/2)$ ($\omega_c$ is the cyclotron frequency, {$\hbar$ is the Planck constant}), the linear dispersion of Dirac electrons results in non-equidistant LL energies, $E_{\pm n}=\pm \hbar\omega_c\sqrt{n}$,
with the LLs of positive ($L_{+n}^s$) and negative ($L_{-n}^s$) energies located symmetrically around 
the zeroth LL, $L_0^s$, with $E_0=0$, here $s=\uparrow,\downarrow$ is for spin state. Both intra- and interband CR transitions are allowed when the orbital indices $n,n'$ of the initial and final LLs satisfy $n'=n\pm1$, resulting in multiple absorption lines centered at different resonant energies 
\begin{equation}\label{eq:DeltaMain}
    \Delta_n^\pm =\hbar \omega_c \left(\sqrt{n+1}\pm\sqrt{n}\right).
\end{equation}

In a proximitized graphene, we find several prominent modifications of the cyclotron absorption. First, the proximity-induced SOC results in the spin-dependent shift of LLs, leading to a polarization-dependent double-peak fine structure of main interband CR lines. Second, CCR transitions (both intra- and interband) between LLs with opposite spin projections are active when the orbital indices of the initial and final LLs satisfy $\Delta n = n'-n \in \{0,\pm 2\}$. 
The corresponding spin-flip absorption is at photon energies different from $\Delta_n^\pm$, which results in the shape deformation of main CRs, and in the appearance of extra CCR absorption lines preceding the main CR interband series.
{The described fine structure of CR is in contrast to previously described features of CR absorption in graphene-based systems~\cite{wakafuji2019detection,ju2020unconventional}}

In our analysis we use a low-energy effective Dirac model augmented by proximity-induced interactions, consistent with  
first-principles studies of proximitized graphene structures~\cite{gmitra2015graphene,gmitra2016graphene,gmitra2017proximity,garcia2018spin,zollner2022engineering,zollner2022proximity,zollner2025first}. 
The unperturbed Dirac Hamiltonian of electrons in graphene 
around the $K,K'$ valleys ($\xi=\pm1$) is~\cite{Katsnelson_2020,castro2009electronic,das2011electronic}
\begin{equation}
\label{eq:H_K}
    H_0 = \hbar v_{\rm F} \bigl(\xi \sigma_x k_x + \sigma_y k_y\bigr), 
\end{equation}
where $v_{\rm F} \approx 1.1 \times 10^8~\mathrm{cm/s}$, $\bm{k}$ is the wave vector counted from $K/K'$, and 
$\bm{\sigma}$ is the vector of Pauli matrices acting on the electron pseudospin (the two inequivalent sublattices $A,B$). 
When proximitized by 
TMDs (e.g., MoS$_2$, WSe$_2$)
or vdW magnetic/AFMs, 
and when the graphene Dirac point lies within the band gap of a neighbouring layer, 
the effective Hamiltonian around $K,K'$ 
acquires extra terms~\cite{gmitra2015graphene,garcia2017SHE,kochan2017model,garcia2018spin,sierra2021van}: 
\begin{equation}
\label{eq:Hprox}
    H_{\rm prox} = H_{\rm vz} + H_{\rm st} +
    H_{\rm R}. 
\end{equation}
Here, $H_{\rm vz} = \xi \hbar \Omega_0s_z/2$ is the valley-Zeeman coupling (opposite sign in $K$ and $K'$), with 
$\bm{s}$ the electron spin Pauli matrix, 
{$\Omega_0$ is an associated Larmor frequency.} 
This term is induced by the large intrinsic SOC of the TMD conduction bands, 
reaching up to tens of meV in W-based compounds and a few meV in Mo-based compounds~\cite{xiao2012coupled,liu2013three,kormanyos2015k}. 
First-principles calculations for graphene/TMD heterostructures find 
$\hbar \Omega_0$ of several meV~\cite{gmitra2015graphene,wang2015strong,garcia2018spin}. 
The term, $H_{\rm st} = U \sigma_z$, 
{is due to on-site asymmetry between the $A,B$ sublattices and its staggered potential, $U$.} 
It opens an energy gap at $k=0$, typically in the $0.2$–$2$~meV range~\cite{gmitra2015graphene,garcia2018spin}. The last term, 
\begin{equation}
\label{eq:H_so}
     H_{\rm R} = \lambda_{\rm R} \bigl(\xi \sigma_x s_y - \sigma_y s_x\bigr),
\end{equation}
is the Rashba coupling due to substrate-induced inversion-symmetry breaking, 
with $\lambda_{\rm R}$ usually in the meV range~\cite{gmitra2015graphene,gmitra2016graphene,zhou2019spin}. 
Importantly, $H_{\rm R}$ involves $\bm{\sigma}$ rather than $\bm{k}$, thus realizing 
the spin–pseudospin coupling (SPC). 
SPC manifests itself in entangled spin–pseudospin dynamics~\cite{tuan2014pseudospin,demoraes2020entanglement}, anomalous polarization of the electric dipole spin resonance~\cite{denisov2024terahertz},
and interband spin pumping~\cite{dugaev2011spin,inglot2014optical}, 
all effects specific to Dirac systems.

In principle, other SOC contributions are also allowed, including the Kane–Mele term 
$\sim \xi \sigma_z s_z$~\cite{kane2005quantum}, 
or a linear-in-momentum Rashba interaction $H_{\rm R}'=\lambda_1 (k_x s_y - k_y s_x)$, both less important than $H_{\rm prox}$. Furthermore, lattice mismatch and finite twist angle can produce moiré structures, 
with proximity-induced radial Rashba effects~\cite{frank2024emergence} and unusual spin–charge conversion~\cite{david2019induced,naimer2021twist,lee2022charge,veneri2022twist,yang2024twist,perkins2024spin}, 
which we do not consider here.

\begin{figure}
    \centering
    \includegraphics[width=1.\linewidth]{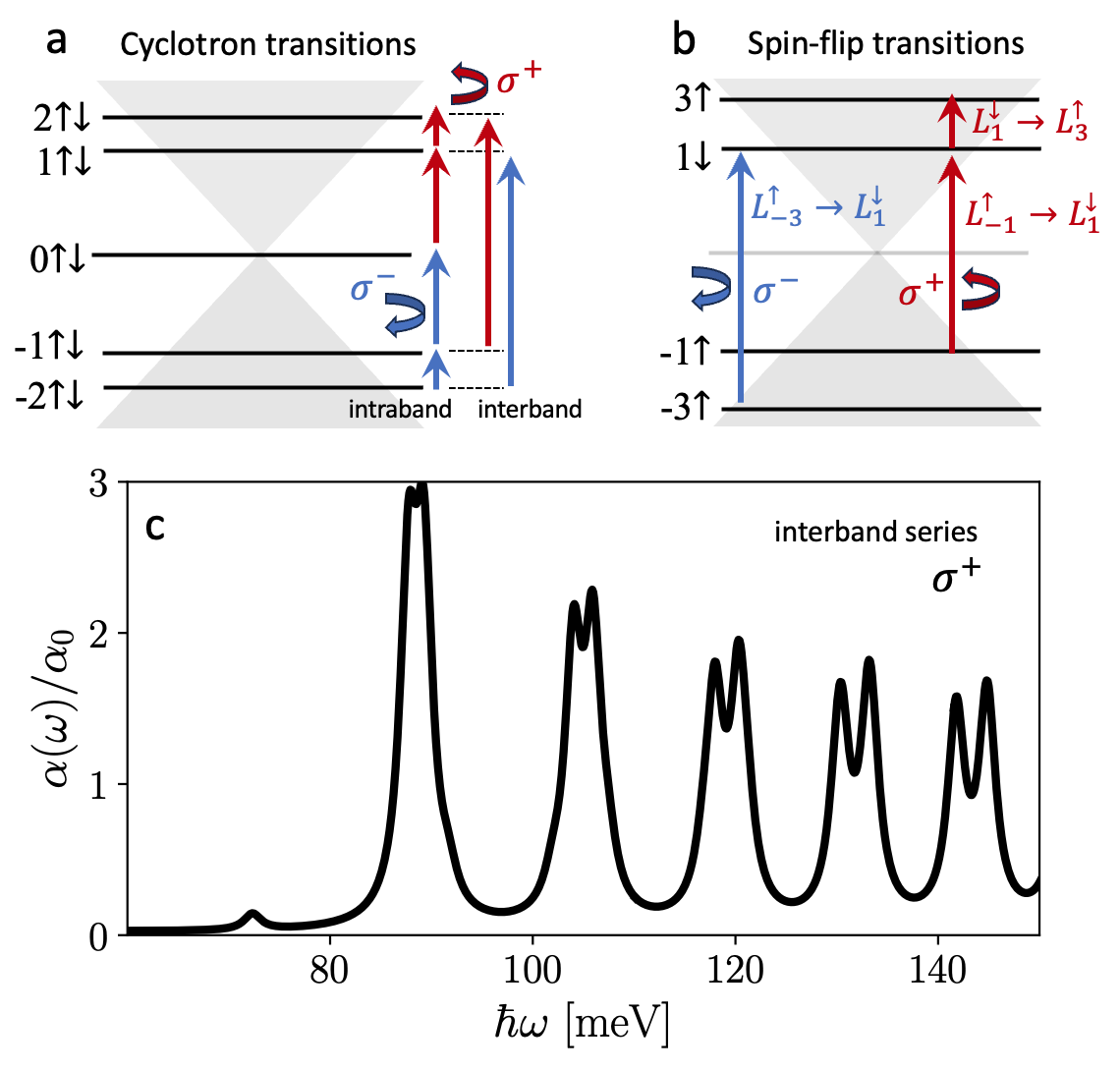}
    \caption{(a) Cyclotron resonance intra- and interband transitions between LLs in a monolayer graphene are active for $\sigma^+$ ($\sigma^-$) polarization when $n\to n+1$ ($n\to n-1$). 
    (b) Combined cyclotron resonance due to spin-flip transitions between LLs with $n\to n\pm(0,2)$. (c) Absorption as a function of $\sigma^+$ photon energy at $B=0.6$~T, the filling factor $\nu=12$, and parameters: $\hbar \Omega_0 =5$~meV, $\lambda_{\rm R}=1.5$~meV, $\Gamma=1$~meV; peaks of the absorption correspond to the interband CR transitions.}
    \label{fig:Scheme}
\end{figure}

We focus on moderate magnetic fields of $0.5$–$2$~T, 
for which the cyclotron energies of low-lying LLs ($\sim25$–$100$~meV) greatly 
exceed all proximity-induced terms (a few meV). This motivates us to start with the LLs of unmodified graphene~\cite{zheng2002hall,brey2006edge,goerbig2011electronic} and 
to discuss the effect of $H_{\rm prox}$ afterwards. After the minimal-coupling substitution 
$\bm{k}\to \bm{k}+(e/c)\bm{A}$ in Eq.~(\eqref{eq:H_K}) 
(with $-e$ the electron charge and $e>0$, by convention, 
$c$ the speed of light), and choosing the Landau gauge $\bm{A}=(0,Bx,0)$ for a static magnetic field $B>0$, we obtain for $K$-valley
\begin{equation}
\label{eq:EWF}
    E_{\lambda n}^0 = 
    \lambda\,\hbar\omega_c \sqrt{n},
    \qquad
    \Psi_{\lambda n s}^0 = \frac{e^{i k_y y}}{\sqrt{2}}
    \begin{pmatrix}
        |n-1\rangle \\[2pt]
        i\lambda\,|n\rangle
    \end{pmatrix} |s\rangle, 
\end{equation}
where $\omega_c = v_{\rm F}\sqrt{2eB/\hbar c}$, 
$\lambda=\pm$ labels 
positive/negative energies, and 
$n\ge 0$ is the orbital LL index. 
Here $k_y$ is a good quantum number in this gauge, 
$|n\rangle$ denotes the $n$th harmonic-oscillator eigenstate with real-space envelope 
$\langle \bm{r}\,|n\rangle=\Phi_n\!\bigl((x-x_0)/\ell_B\bigr)$, 
centered at $x_0=\ell_B^2 k_y$, and $\ell_B=\sqrt{\hbar c/(eB)}$; 
the spinor part is $|s\rangle=|\uparrow,\downarrow\rangle$. 
For $K'$ one may take $\sigma_x\Psi_{\lambda n s}^0$ (up to an overall phase).  
The case $n=0$ is special: 
the zeroth LL has exactly zero energy and 
exhibits full sublattice polarization (on $B$ for $K$ and on $A$ for $K'$). 
We also note the non-equidistant positions of the LL energies, $E_{\pm n}^0 \propto \pm\sqrt{nB}$.

The optical transitions between LLs 
due to an electric field, $\bm{E}_\omega^\eta e^{- i \omega t}$,  
with polarization $\bm{E}_\omega^\eta=E_\omega \bm{e}_\eta$  
are derived in EDA 
from the Hamiltonian, 
$\mathcal{V} = ({ie}/{2\omega}) 
(\bm{v E}_\omega^\eta)$,
with the velocity operator $\bm{v}=v_{\rm F} \bm{\sigma}$, and $v^\pm=v_F\sigma_\pm$
for the $\eta=\pm$ circular polarizations. 
The matrix element  
$v_{nn'}^\pm$ between $\Psi_{\lambda ns}^0$ and $\Psi_{\lambda'n's}^0$ 
is spin-independent with 
$v_{nn'}^\pm =a v_{\rm F} \delta_{n,n'\pm1}$, 
here $a=1/2$ for $n,n'\neq0$, otherwise  $a=1/\sqrt{2}$. 
The orbital selection rules of main CRs are encoded in $v_{nn'}^\pm \propto \delta_{n,n'\pm1}$: 
$\sigma^\pm$ light leads to the transitions with the increase (decrease) of the LL index $n$ by one~\cite{gusynin2007anomalous} 
(for all $\xi,s$), 
including intra- ($\lambda=\lambda'$) and 
interband ($\lambda=-\lambda'$) processes. 
Different CR transitions with low-lying LLs for $\sigma^\pm$
are shown in Fig.~\ref{fig:Scheme}a.

Let us comment on accounting for proximity effects from Eq.~(\ref{eq:Hprox}) for graphene LLs. 
First, $H_{\rm vz}$ does not couple to orbital degrees of freedom and simply produces a valley-dependent Zeeman spin splitting, $E_{\lambda n s}= E_{\lambda n}^0 + \xi \hbar \Omega_0 s$, 
leaving $\Psi_{\lambda n s}^0$ unchanged 
but setting the spin quantization axis along $\hat{\bm{z}}$ ($s=\pm{1}/{2}$). 
The staggered potential $H_{\rm st}=U\sigma_z$ 
is spin independent: it 
couples LLs with the same orbital index $n$ 
but opposite band index $\lambda$ via
$\sigma_z \Psi_{\lambda n s}^0=\Psi_{-\lambda n s}^{0}$. 
For $n\neq 0$, 
this yields an energy correction of order $(U/
\Delta_n^+)^2$,
which is negligible compared to $\hbar \Omega_0$. 
The special case $n=0$ is different: 
the LL is fully sublattice polarized and is an eigenstate of $\sigma_z$, giving a linear-in-$U$ shift 
$E_{0 s} = \xi \bigl(\hbar \Omega_0 s - U\bigr)$. 
Thus, the only effect of $H_{\rm st}$ we retain is this $n=0$ shift. 
Including $H_{\rm vz}+H_{\rm st}$ 
does not couple spin and orbital parts 
of the LL wave functions and, consequently, 
does not induce spin-flip transitions 
and fine structure of CRs.

To capture the SOC-driven change in the absorption, we include $H_{\rm R}$, 
which can be written as ($K$ valley):
\begin{equation}
    \label{eq:HRashbaUP}
    H_{\rm R} = \frac{i\lambda_{\rm R}}{2v_{\rm F}}\!\left( 
    v^{+} s_- - v^{-} s_+
\right),
\end{equation}
with $s_\pm = s_x \pm i s_y$. 
This form shows that $L_n^{s}$ 
mixes with $L_{n\pm1}^{-s}$ 
having opposite spin, $-s$, and orbital index, $n\pm1$, 
as implied by $v_{n n'}^{\pm} \propto \delta_{n,n'\pm 1}$. In our analysis, we treat $H_R$ in the LLs basis of Eq.~(\ref{eq:EWF}) with $E_{\lambda n s}= E_{\lambda n}^0 + \xi \hbar \Omega_0 s - \xi \delta_{n,0}U$ and perform the exact numerical diagonalization of a truncated system with total number of LLs $N_L = 30$. Low-lying LLs in graphene with multiple SOC terms have been considered in~\cite{frank2020landau}. 
Note that the SOC-induced 
orbital mixing can be also seen in excitonic effects with account for a finite Coulomb interaction~\cite{cao2025tunable}.

We proceed with analyzing the absorption coefficient, $\alpha_\eta(\hbar\omega)$, 
for $\bm{E}_\omega^\eta=E_\omega \bm{e}_\eta$, 
including $\sigma^\pm$ circular polarizations with $\bm{e}_\pm=(1,\pm i)/\sqrt{2}$. 
We make use of the Kubo–Greenwood formula~\cite{orlita2010dirac} for $\alpha_\eta$:
\begin{equation}
    \alpha_\eta(\hbar \omega) = 
    \frac{\alpha_0 \hbar \omega_c^2}{\omega v_F^2}
    \sum_{r,r',\xi}|v_{rr'}^\eta|^2 (f_r-f_{r'})  
    \delta_\Gamma(\hbar\omega+E_{r'}-E_r),
\end{equation}
where $v_{rr'}^\eta$ 
is the matrix element of the velocity operator, $v_F\sigma_\eta$, between 
numerically obtained eigenstates $\Psi_{\lambda n s}$ (indices $r=(n,s)$ and $r'=(n',s')$), 
$\alpha_0= \pi e^2/\hbar c \approx 0.023$ 
is the universal absorption of graphene, 
$f_r$ is the Fermi-Dirac function, 
and $\delta_\Gamma(x) = (\Gamma/\pi)(x^2+\Gamma^2)^{-1}$ is the Lorentzian of finite width $\Gamma\approx1$~meV~\cite{orlita2011carrier}.

\begin{figure}[t]
    \centering
    \includegraphics[width=1.\linewidth]{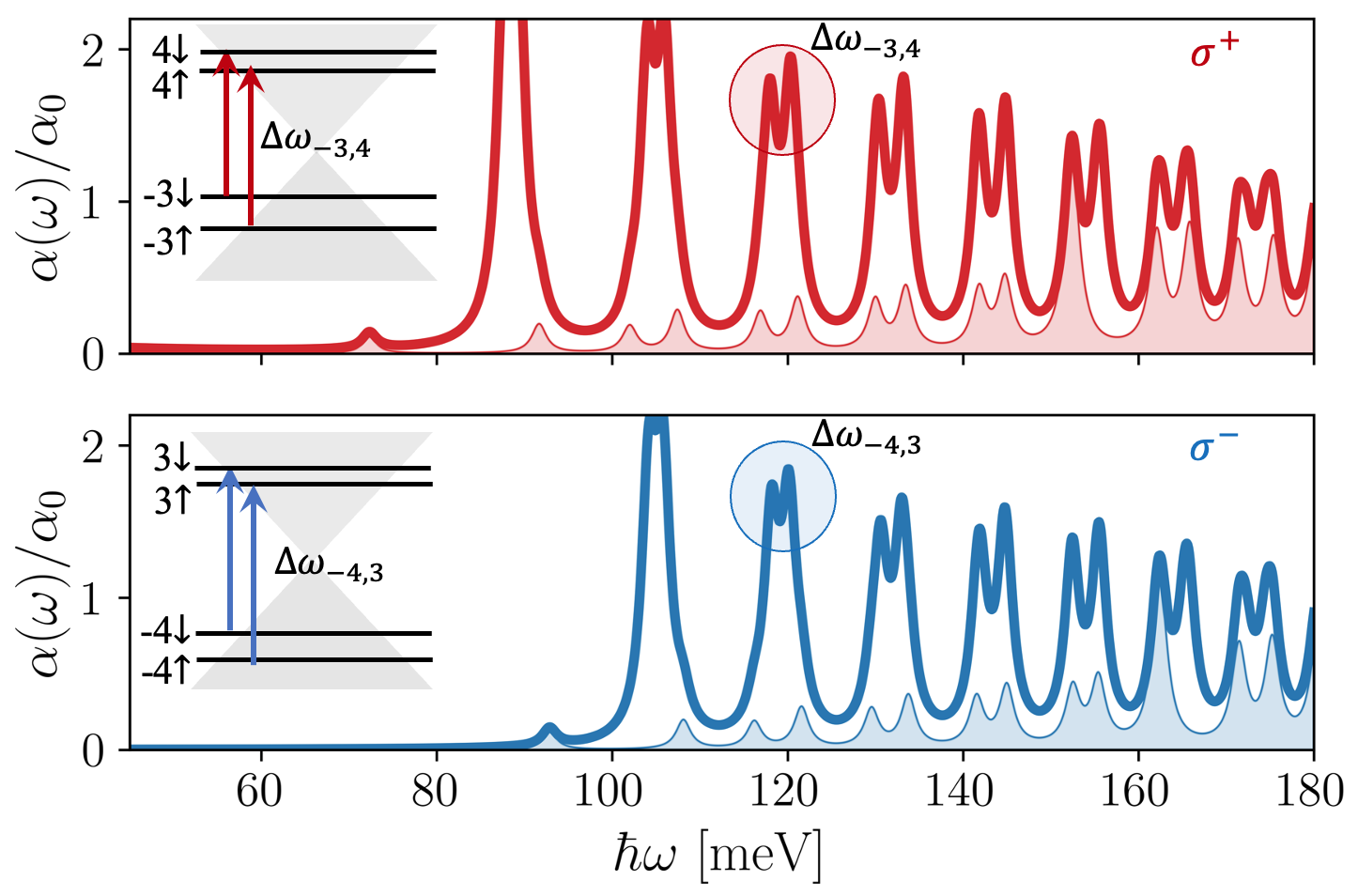}
    \caption{Absorption $\alpha_\pm(\hbar \omega)$ for $\sigma^+$ (a) and $\sigma^-$ (b) polarizations; the shaded regions show a partial contribution due to spin-flip transitions allowed in the presence of $H_{\rm R}$. 
    The inset in (a) [(b)] shows the frequency difference for $L_{-3}^s\to L_4^s$ [$L_{-4}^s\to L_3^s$] 
    responsible for a double peak fine structure around 
    $\hbar \omega_{-3,4}$ [$\hbar \omega_{-4,3}$]. 
    The parameters: $\nu=12$, $B=0.6$~T, $\hbar \Omega_0 = 5$~meV, $\lambda_{\rm R}=1.5$~meV, $\Gamma=1$~meV. }
    \label{fig:ft2}
\end{figure}

The full absorption $\alpha_\pm(\hbar\omega)$ 
is shown in Fig.~\ref{fig:Scheme}c and Fig.~\ref{fig:ft2} 
for $B=0.6$~T and Fermi energy, $\mu=40$~meV, 
corresponding to the filling factor $\nu=12$ 
[$L_3^{\uparrow,\downarrow}$ are the first unoccupied LLs]
{at temperature, $T\ll \hbar \Omega_0$.}
The absorption is strongly dominated by the main CRs of graphene, with the resonant frequencies for intra- and interband transitions centered at $\Delta_n^\mp$ from Eq.~(\ref{eq:DeltaMain}); 
Fig.~\ref{fig:Scheme}c and Fig.~\ref{fig:ft2} capture frequency range of the interband series only. 
The proximity-induced Rashba SOC manifests itself in several features of the absorption spectrum: (i) a double-peak fine structure of main CR lines (ii) appearance of additional lines with smaller magnitude and shoulders of main CRs 
due to spin-flip absorption.

For (i), we note that $H_{\rm R}$
results in the mixing of LLs with opposite spin states and orbital indices differing by one. This mixing gets stronger for states with larger orbital index due to the concentration of LL energies ($\Delta_n^-\propto\sqrt{n+1}-\sqrt{n}$)
and, eventually, leads to a larger energy shift $\propto (\lambda_{\rm R}/\Delta_n^-)^2$ of LLs. Importantly, the inclusion of the valley Zeeman splitting, $H_{\rm vz}$, also modifies the exact energy difference between $L_n^s$ and $L_{n\pm 1}^s$, 
such that the actual energy shift of $L_n^s$ due to the Rashba SOC follows $\lambda_{\rm R}^2/(\Delta_n^- +s \Delta_{{\rm vz}})^2$ and becomes spin-dependent. 
This leads to a finite attenuation of the resonant frequencies of the main CR transitions between the LLs with same orbital indices and opposite spin states (e.g. $L_{-n}^\uparrow \to L_{n+1}^\uparrow$ and $L_{-n}^\downarrow \to L_{n+1}^\downarrow$), which results in a double-peak 
structure in Figs.\ref{fig:Scheme},\ref{fig:ft2}
for the interband absorption. 
{This fine structure evolves 
with $T$ due to a thermal broadening of $\Gamma$ 
and LLs occupancies, 
as well as it changes with $\lambda_{\rm R}$, 
see the Supplemental Material~\cite{SM}.}

For $\sigma^+$ from Fig.~\ref{fig:ft2}a and $L_{-3}^s\to L_4^s$ transitions the unperturbed resonant energy is $\hbar \omega_{-3,4}^0\approx120$~meV. The fine structure of this line, $\Delta \omega_{-3,4}$ [see the inset in Fig.~\ref{fig:ft2}a], appears mostly due to the spin-dependent shift of $E_{4\uparrow}-E_{4\downarrow}\propto \lambda_{\rm R}^2 \Delta_{\rm vz} / \Delta_4^3$.
For $\sigma^-$ from Fig.~\ref{fig:ft2}b and $L_{-4}^s\to L_3^s$ transitions, centered at the same resonant energy $\hbar \omega_{-4,3}^0\approx120$~meV,
the fine structure, 
$\Delta \omega_{-4,3}$ [see the inset in Fig.~\ref{fig:ft2}b],
is mostly due to $E_{-4\uparrow}-E_{-4\downarrow}\neq 0$.

The fine structure is realized for all CR interband transitions and gets stronger at larger $n$. Ultimately, with $n\gg 1$, the notion of a well-defined spin quantization axis for a single LL becomes invalid: The Rashba SOC results in a strong mixing of LLs with opposite spin states, hence producing a spin-textured wave function. In this regime, there is no well-defined spin selection rule for CR transitions, and multiple absorption lines appear in $\alpha_\pm$. In particular, the shaded regions in Fig.~\ref{fig:ft2} display the partial contributions to the full absorption due to transitions between LLs with opposite spin states in $\lambda_{\rm R}\to 0$ limit. At large $n$, each double peak structure has a noticeable contribution from the so-defined spin-flip transitions.

Let us now focus more closely on (ii): CCR and its spin-flip transitions. 
It is instructive to consider the linear in $H_R$ corrections to the wave functions:
\begin{equation}
    \label{eq:EqWFpert}
    \Psi_{\lambda n s}=\Psi_{\lambda n s}^0 + \delta \Psi_{\lambda n s},
\quad
\delta \Psi_{\lambda n s} = 
\frac{i \lambda_{\rm R}}{2 v_{\rm F}} \sum_{\lambda' n'} C_{n n'}^{s s'} \,\Psi_{\lambda' n' s'}^{0},
\end{equation}
where $s'=-s$, and 
\[
    C_{n n'}^{\uparrow \downarrow}
    = \frac{v_{n' n}^{+}}{E_{\lambda n \uparrow} - E_{\lambda' n' \downarrow}},
    \qquad
    C_{n n'}^{\downarrow \uparrow}=
    \frac{-\,v_{n' n}^{-}}{E_{\lambda n \downarrow } - E_{\lambda' n' \uparrow}}. 
\]
As an example, 
the $L_1^{\uparrow}$ state 
acquires intra- and interband admixtures from $L_{0,2}^{\downarrow}$ 
and $L_{-2}^{\downarrow}$, respectively.
In the EDA, the matrix element of the velocity operator, $v_{rr'}^\pm$  
corresponding to the spin-flip transition from $\Psi_{\lambda' n's'}$ to $\Psi_{\lambda n s}$ can be written as 
\begin{equation}
\label{eq:V}
    v_{rr'}^\pm
    = \frac{i \lambda_{\rm R}}{2 v_{\rm F}}  \sum_{m} 
    \Big(   v_{n m}^{\pm} C_{m n'}^{s s'}
    + C_{n m}^{s s'} v_{m n'}^{\pm} \Big),
\end{equation}
where the sum runs over intermediate states with orbital index $m$. 
Since both $C_{n m}^{s s'}\propto \delta_{n,\,m \pm 1}$
and $v_{m n'}^\pm \propto \delta_{m,\,n'\pm 1}$
are nonzero only when $n$ and $m$ differ by $\pm 1$, the resulting selection rules for CCR require that the
orbital indices $n,n'$ either coincide 
or differ by $\pm 2$: $n' = n \pm (0,2)$. 
The derived selection rules 
for the CCR are identical in the $K$ and $K'$ valleys and are summarized in Table~\ref{tab:transitions}. 
As an example, in Fig.~\ref{fig:Scheme}b we show CCR transitions involving  $L_1^\downarrow$ in the $K$ valley:
there are two interband transitions from $L_{-1}^\uparrow$ and $L_{-3}^\uparrow$, 
and a single intraband transition to $L_3^\uparrow$, all three in the IR range. 
If $L_1^\uparrow$ is not fully occupied, there is also a spin-flip transition $L_1^\downarrow \to L_1^\uparrow$ with energy $\hbar \omega \approx \hbar \Omega_0$ in the THz range. 
The derived rules are also reproduced in our numerical calculations based on the diagonalization of truncated LLs series.

\begin{table}[t]
    \centering
    \label{tab:transitions}
    \begin{tabular}{|c|c|c|}
        \hline
        & CR & combined spin-flip \\
        \hline
        $\sigma^+$ & $n \to n + 1$ & $n\uparrow \to n\downarrow$ \quad $n\downarrow \to n + 2\,\uparrow$ \\
        \hline
        $\sigma^-$ & $n \to n - 1$ & $n\downarrow \to n\uparrow$ \quad $n\uparrow \to n - 2\,\downarrow$ \\
        \hline
    \end{tabular}
    \caption{Selection rules for the cyclotron resonance (CR) and combined spin-flip cyclotron resonance (CCR).   }
\end{table}

\begin{figure}[b]
    \centering
    \includegraphics[width=1.\linewidth]{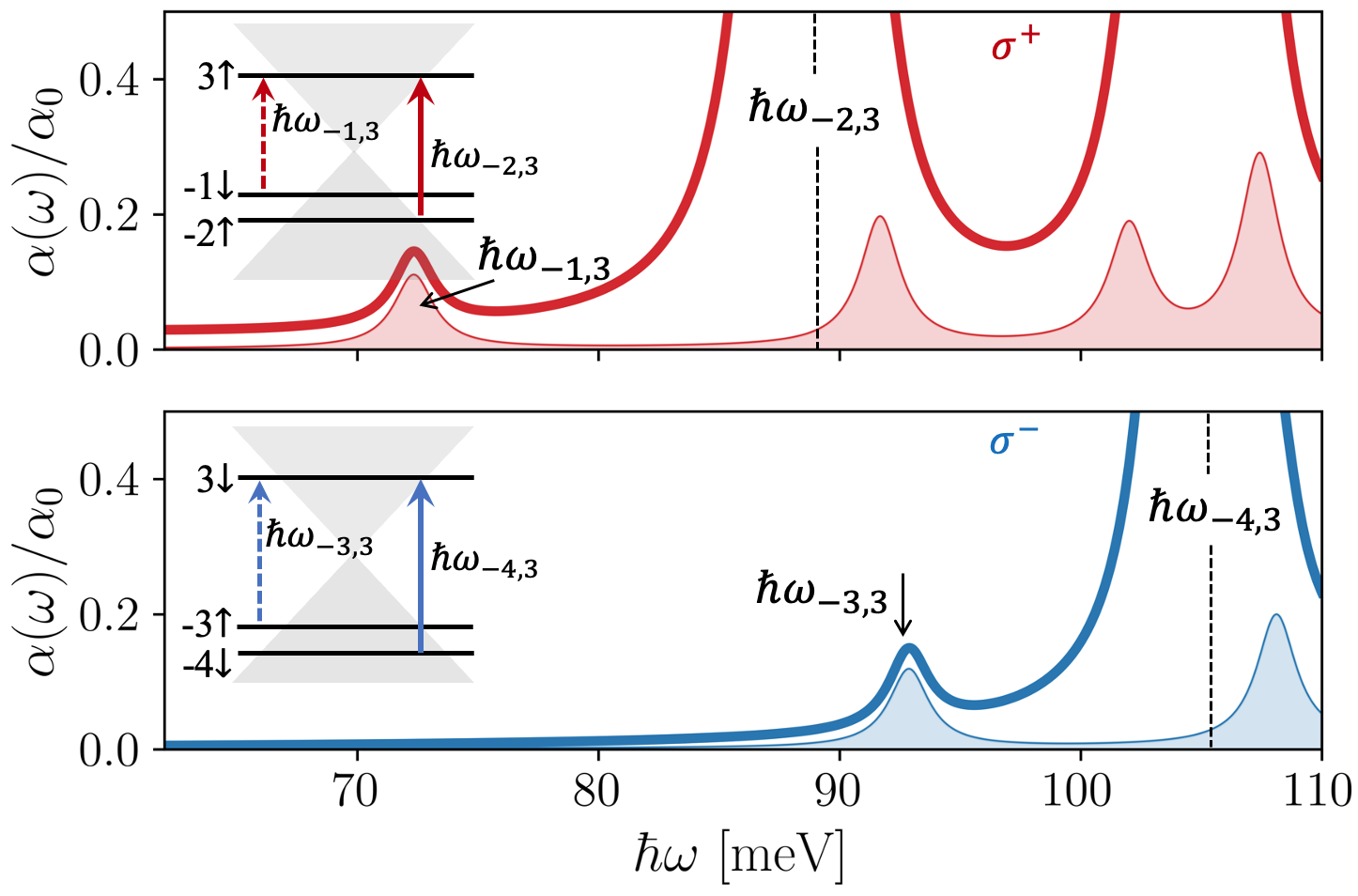}
    \caption{Absorption $\alpha_\pm(\hbar \omega)$ for $\sigma^+$ (a) and $\sigma^-$ (b) polarizations. The shaded regions show the CCR contribution due to $H_{\rm R}$-induced spin-flip absorption. At $\nu = 12$, there is a well-resolved CCR line $L_{-1}^\downarrow\to L_3^\uparrow$ for (a) and to 
    $L_{-3}^\uparrow\to L_3^\downarrow$ for (b). The corresponding CCR transitions are shown in insets.}
    \label{fig:ft1}
\end{figure}

For small $n$ and large $\Delta_n^-$ (away from a strong mixing of LLs with opposite spin states),
the spin-flip absorption of CCRs is centered at the resonant 
energies well separated from the main CR lines ($\Delta_n^\pm$): 
the interband CCR transitions are at $2\hbar\omega_c\sqrt{n}$ and $\hbar\omega_c\!\left(\sqrt{n+2}+\sqrt{n}\right)$,
while intraband CCRs are at $\hbar\omega_c\!\left(\sqrt{n+2}-\sqrt{n}\right)$. 
Herewith, the magnitude of CCR contribution to the absorption, $\delta\alpha_{\rm R} \propto (\lambda_{\rm R}/\Delta_n^-)^2$, is relatively weak due to small ratio of $\lambda_{\rm R}/\Delta_n^-$, which results in non-Lorentzian profile of CR absorption lines, visible in Figs.~\ref{fig:Scheme},~\ref{fig:ft2} for a couple of transitions in the beginning of interband series. As we discussed above, at larger $n$ [and smaller $\Delta_n^-$], the Rashba-induced mixing gets stronger and modifies $\alpha_\pm$ in a stronger way.

Importantly, there are certain filling factors that allow us to isolate a single 
CCR absorption line (for both $\sigma^\pm$), see Fig.~\ref{fig:ft1}. 
As a general trend, the magnitude of $\delta \alpha_{\rm R} \propto (\lambda_{\rm R}/\Delta_n^-)^2$ increases for larger LLs indices
due to decrease in $\Delta_n^-\propto (\sqrt{n+1}-\sqrt{n})$. 
Hence, to have a stronger CCR absorption we should focus on finite $\nu$. 
At the same moment, to achieve better contrast for CCRs, 
we note that at larger $\nu$, intra- and interband transitions 
(both for CRs and CCRs) get separated in energies, so it would be preferable to have a spin-flip resonant line lying 
within this energy gap 
where the main CR absorption is suppressed the most. 
In fact, the CCR selection rules 
provide us with such a configuration at 
$\nu = 4 \times l$, where $l$ is an integer: In this case we have a well-resolved CCR transition precursoring the interband CRs for both $\sigma^\pm$.

This is illustrated in Fig.~\ref{fig:ft1}  
for the case of 
$\nu=4 \times 3=12$, when three lowest LLs are filled out with electrons ($B=0.6$~T). 
In case of $\sigma^+$ from Fig.~\ref{fig:ft1}a,
the highest energy of the intraband CR ($\hbar \omega_{2,3}^- = 25$~meV for $L_{2}^s\to L_3^s$) is well below the lowest energy of the interband CR ($\hbar \omega_{-2,3}^+ \approx 90$~meV, for $L_{-2}^s\to L_3^s$). At the same moment, there is a CCR spin-flip transition $L_{-1}^\downarrow\to L_3^\uparrow$ with frequency $\hbar \omega_{-1,3} \approx 74$~meV, below 
$\hbar \omega_{-2,3}^+$. 
Similarly, in case of $\sigma^-$ polarization in Fig.~\ref{fig:ft1}b, 
the CCR transition $L_{-3}^\downarrow\to L_3^\uparrow$ has its energy, $\hbar \omega_{-3,3} \approx 94$~meV, smaller than the first main interband CR line of $\hbar \omega_{-4,3} \approx 106$~meV.

These two CCR peaks (for $\sigma^\pm$) are both related to the interband spin-flip absorption and appear specifically due to the spin-pseudospin type of the Rashba coupling. 
Using the minimal coupling substitution, $\bm{k}\to\bm{k}+(e/c)\bm{A}$, 
in the $k$-linear Rashba term, $H_R'$, we get the spin-light interaction, 
$H_{\rm int}'=(e\lambda_1/c) (A_x s_y - A_y s_x)$, 
that does not act on the orbital part of 
LL wave functions. 
Hence, in the leading order by $\lambda_1$, 
$H_{\rm int}'$ leads only to spin-flip intraband transitions with unchanged orbital indices, 
$n'=n$. 
Detecting finite spin-flip interband absorption lines, as from Fig.~\ref{fig:ft1}, 
could be a direct confirmation of SPC over the 
k-linear Rashba interaction.

We finally comment on the polarization dependence 
of the absorption. 
The fine structure of interband CR transitions and 
the extra CCR precursor lines are present for both $\sigma^\pm$. 
A finite spin-flip absorption active in both $\sigma^\pm$ is in contrast to the electron paramagnetic resonance due to magneto-dipole transition~\cite{slichter2013principles} 
active only for a single circular polarization. 
Our situation, however, is similar to the intraband spin-flip electric dipole spin resonance (EDSR) of graphene with the spin-pseudospin Rashba coupling and in the absence of Landau levels~\cite{denisov2024terahertz}. 
In the latter case, the EDSR can be understood based on the coupled spin-pseudospin resonance dynamics~\cite{denisov2024terahertz} with the spin torque determined by the symmetry of SOC rather than by the polarization of electric field.
Finally, a finite difference in the absorption for $\sigma^\pm$ leads to the magneto-optical response of media and a nonzero Kerr effect (MOKE). 
Our calculations suggest that a finite MOKE signal can be induced in a proximitized graphene around the peaks of the interband CR absorption. 
For the ordinary CR of a pristine graphene, 
the transitions $L_{-n}^s\to L_{n+1}^s$
and $L_{-n-1}^s\to L_{n}^s$ are centered at the same frequencies and are induced in $\sigma^+$ and $\sigma^-$, respectively, with the same oscillator strength, hence no MOKE. 
The inclusion of finite Rashba and valley-Zeeman SOC terms results in the double-peak structure of interband CR absorption lines which, in addition, is polarization dependent, see Fig.~\ref{fig:ft2} for $\sigma^\pm$ absorption panels. This indicates a finite MOKE for $\hbar \omega$ around the peaks of interband CR lines.
We also note that our MOKE can be further modified by including the valley asymmetry proximity mechanisms~\cite{choi2023asymmetric,zhou2023asymmetry}.

To summarize, our findings suggest that the 
the cyclotron resonance spectroscopy, its mid-infrared range absorption and the associated MOKE responses are useful for probing the proximity-induced spin interactions in graphene/TMDs heterostructures.

\section*{Data Availability}
There are no publicly available research data or software supporting this manuscript. Requests for further information or data should be sent to the authors.

\section*{Acknowledgements}
We thank Igor Rozhansky and Igor {\v{Z}}uti{\'c} for valuable discussions.  
Analytical and numerical calculations of the cyclotron resonance absorption coefficient were carried out with support from the Russian Science Foundation under Grant No.~25-12-0093. 
\bibliography{Ref}

\end{document}